\documentclass[doublecol]{epl2}

\usepackage[T1]{fontenc}
\usepackage[latin1]{inputenc}
\usepackage{amsmath,amsfonts}
\usepackage{graphics,color}
\usepackage[english]{babel}

\newcommand{\bfp}{{\mathbf{p}}}

\newcommand{\bfr}{{\mathbf{r}}}
\newcommand{\bfv}{{\mathbf{v}}}
\newcommand{\bfn}{{\mathbf{\hat n}}}
\newcommand{\pdp}{{\mathbf{p}\cdot\delta\mathbf{p}}}
\newcommand{\asin}{\operatorname{asin}}

\def\surname#1{#1}

\begin{document}

\title{Self-propelled hard disks: implicit alignment and transition to collective motion}

\author{Khanh-Dang \surname{Nguyen Thu Lam}, Michael \surname{Schindler},
        Olivier \surname{Dauchot}}

\institute{UMR Gulliver 7083 CNRS, ESPCI ParisTech, PSL Research University, 
            10~rue Vauquelin, 75005~Paris, France}


\date{\today}%

\abstract{%
We show that low density homogeneous phases of self propelled hard disks
exhibit a  transition from isotropic to polar collective motion, albeit of a
qualitatively distinct class from the Vicsek one. In the absence of noise, an
abrupt discontinuous transition takes place between the isotropic phase and a
fully polar absorbing state. Increasing the noise, the transition becomes
continuous at a tri-critical point. We explain all our numerical findings in
the framework of Boltzmann theory, on the basis of the binary scattering
properties. We show that the qualitative differences observed between the
present and the Vicsek model at the level of their phase behavior, take their
origin in the complete opposite physics taking place during scattering events.
We argue that such differences will generically hold for systems of
self-propelled particles with repulsive short range interactions.
}%
\maketitle

Self-propelled particles borrow energy from their environment and
convert it to translational motion. Depending on their density and
interaction details, energy taken up on the microscopic scale can be
converted into macroscopic, collective motion. A transition from an
isotropic phase to a polar, collective phase is observed. Most
theoretical knowledge about such \emph{active matter} and its
transitions has been developed for Vicsek-like models: point particles
travel at constant speed, to represent self-propulsion, and their
interaction rules comprise both explicit alignment and noise, to account
for external or internal perturbations.
In addition, self-diffusion noise may be incorporated in the individual
motion. Typically, the alignment strength and noise intensity control
the transition between the isotropic and the polar phases~\cite{
Vicsek1995, Toner1995, Bertin2006, Chate2008, Ihle2011, Peshkov2014
}.

In order to describe more closely experiments with
bacteria~\cite{Zhang2009, Zhang2010, Chen2012} or with self-propelled actin
filaments~\cite{ Schaller2010 }, other models incorporates the
alignment mechanism through the steric interaction of elongated
objects~\cite{ Peruani2006, Baskaran2008, Ginelli2010 }. The interaction
accounts for the body-fixed axis of these particles, which is in some sense
more \emph{microscopic} than the purely effective (mesoscopic) alignment
rules in Vicsek-like models.  Steric interaction is not the only
microscopic interaction that can lead to collective behaviour. An
alternative way to use dynamics for aligning (hard or soft) circular disks is
to incorporate inelasticity or softness in the interaction rules~\cite{
Aranson2005, Grossman2008, Henkes2011, Hanke2013}. Not surprisingly,
also these models exhibit a transition to a globally aligned phase.

A model experimental system of self-propelled disks has been realized by
vertically vibrating millimeter-sized circular metal disks with two
different ``feet''~\cite{Deseigne2010, Deseigne2012}. When isolated, such
particles advance in the direction of their body axis, with a reasonably
high persistence length. When sufficiently crowded, they start to align to
each other and move along in dense clusters.  Successful modelling of this
experiment, in the form of Newton's equations for rigid disks with a force
acting on the center and a torque turning it around, plus slightly
inelastic collisions and some active noise included, have demonstrated that
purely repulsive mechanical interaction with no explicit alignment, were
sufficient ingredients to observe a transition to collective motion~\cite{
Weber2013 }. The nature of this effective alignment remains however
mysterious. Is inelasticity a required ingredient? Or, as suggested
in~\cite{ Deseigne2010, Deseigne2012, Weber2013 }, does it come from
repeated recollisions of particle pairs, during which the velocity vector
repeatedly converges against the body axis, but also the body axis has time
to relax towards the time-average of the velocity axis.  Given that the
experiment is very much the one closest to a simple hard-disk liquid and
lends itself to a description in terms of statistical physics,
understanding precisely its phase transitions, the effective alignment
mechanism, and the role of the noise would be a major step towards a
theoretical description of active liquids in general.%
\begin{figure}
  \centering
  \includegraphics{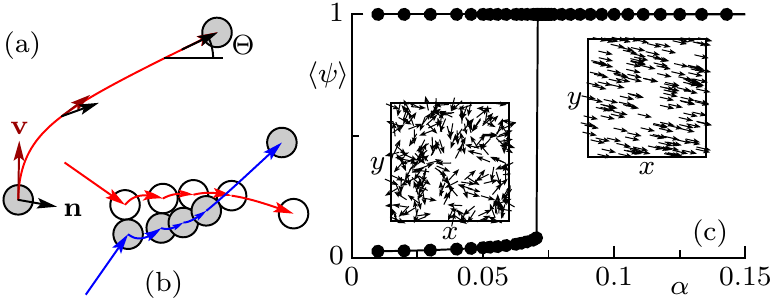}%
  \caption{\label{fig:1}%
  \textbf{(a)}~An isolated self-propelled particle converges to its stationary
  state where velocity~$\bfv$ and polarity~$\bfn$ are parallel. \textbf{(b)}~A
  single binary ``scattering event'' can consist of many hard-disk elastic
  collisions. \textbf{(c)}~Stable phases in the absence of noise. Between
  isotropic ($\langle\psi\rangle\approx0$) and polar ($\langle\psi\rangle=1$)
  phase is a discontinuous transition.}
\end{figure}

In this letter we provide such an understanding, restricting ourselves to
the analysis of spatially homogeneous phases in the low density limit. To do so
we proceed in three steps. (i)~We perform molecular dynamics of the model
equations for purely elastic interactions, with and without noise: In the
absence of noise, the system exhibits a strongly first order transition from
the isotropic to the collective motion phase (see Fig.~\ref{fig:1}c). Above a
finite level of noise, the transition becomes second order -- a tricritical
point exists. This establishes the phase behaviour which we will explain
from theoretical considerations. (ii)~We analyze the model equations on the
grounds of the Boltzmann equation, by making use of a recently proposed
observable~$\langle\pdp\rangle$ which quantifies the
non-conservation of momentum~\cite{pdp}.
This observable allows to span the bridge from the microscopic
dynamics, in particular binary collisions such as depicted in
Fig.~\ref{fig:1}b, to macroscopic order parameters. From a direct
sampling of all possible binary scattering events, we obtain an excellent
quantitative prediction of our numerical findings. (iii)~We scrutinize the very
peculiar dynamics of a collision between two self propelled disks and explain
the specific shape of the scattering function that was obtained numerically
in~(ii). We further find that recollisions are not necessary for the observed
alignment, contrary to our previous belief.

\subsection{Model of self-propelled hard disks}
The model consists of $N$~hard disks in a square box of size~$L\times
L$, with periodic boundary conditions.  The density is~$\rho=N/L^2$.
Particles, being self-propelled, relax to a stationary speed~$v_0$. As
units of length and time we choose the diameter~$d_0$ of the particles
and~$d_0/v_0$, respectively. A particle~$i$ has coordinates~$\bfr_i$,
velocity~$\bfv_i$, and a body axis given by the unit vector~$\bfn_i$
(see Fig.~\ref{fig:1}a). Between collisions, it evolves according to the
equations
\begin{subequations}
\label{eq:dyn}
\begin{align}
  \label{eq:dotbfr}
  \tfrac{\text{d}}{\text{d}t}\bfr_i &= \bfv_i , \\
  \label{eq:dotbfv}
  \tau_v
  \tfrac{\text{d}}{\text{d}t}\bfv_i &= \bfn_i - \bfv_i,\\
  \label{eq:dotbfn}
  \tau_n
  \tfrac{\text{d}}{\text{d}t}\bfn_i &= (\bfn_i \times \hat\bfv_i) \times \bfn_i .
\end{align}
\end{subequations}
The competition between the self-propulsion~$\bfn$ and the viscous
damping $-\bfv$ in Eq.~\eqref{eq:dotbfv} lets the velocity relax
to~$\bfn$ on a timescale~$\tau_v$. Similarly, in Eq.~\eqref{eq:dotbfn},
the polarity $\bfn$ undergoes an overdamped torque that orients it
toward~$\bfv$ on a timescale~$\tau_n$. Interactions between particles
are elastic hard-disk collisions which change $\bfv$ but not~$\bfn$.
After such a collision, $\bfv$~and $\bfn$ are not collinear, and the
particles undergo curved trajectories which are either interrupted by
another collision (Fig.~\ref{fig:1}b), or the particles reach their
stationary state, where $\bfv=\bfn$ and the trajectory is straight at a
speed $v_0=1$ (Fig.~\ref{fig:1}a). The final direction of $\bfv$ (equal
to that of~$\bfn$) depends on the parameter
\begin{equation}
  \alpha=\tau_n/\tau_v ,
\end{equation}
which can be understood as the persistence of the polarity~$\bfn$.
Linearizing the evolution equations around the stationary state, one can
show that the final polar angle is given by the weighted average of the
initial angles, $(\theta_n + \alpha \theta_v)/(1+\alpha)$. When
$\alpha\ll1$, $\bfn$~is practically always directed along~$\bfv$.

On top of the deterministic trajectories given by the
Eqs.~\eqref{eq:dyn}, we add some angular noise by the following
procedure. Given a time step~$\delta t\to0$, we rotate $\bfv_i$~and
$\bfn_i$ by the same angle~$\eta_i(t)$, distributed normally with zero
mean and variance~$2D\delta t$, where the constant $D\ge0$ fixes the
level of the angular noise. Noises of different particles are
statistically independent. We choose $\delta t$ much smaller than all
other timescales in the dynamics. The relevant parameter to characterize
the angular noise is then~$D/\lambda$, where $\lambda=4\rho/\pi$ is the
characteristic scattering rate of the system, which is proportional to
the density~\cite{pdp}.

%
\begin{figure}
  \centering
  \includegraphics{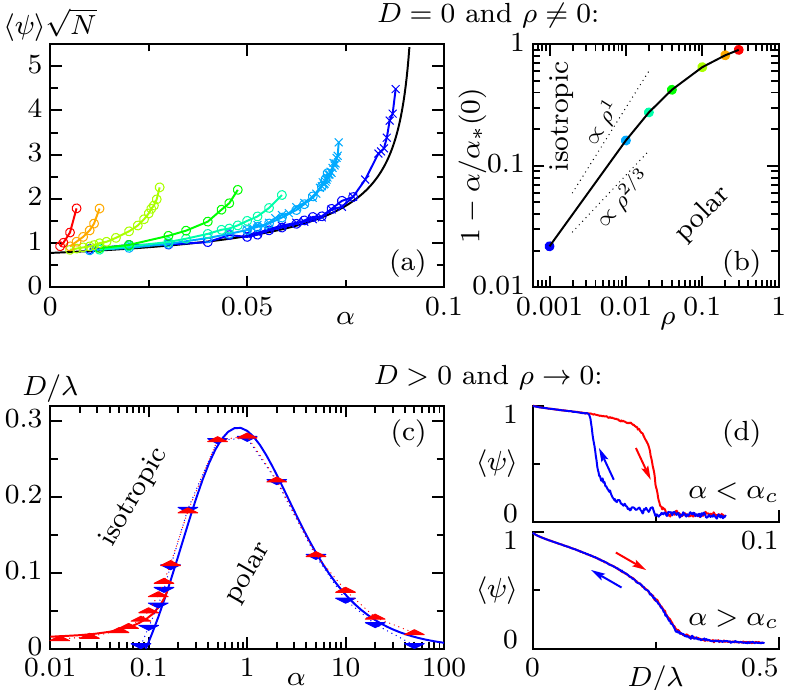}%
  \caption{\label{fig:transition}%
  \textbf{(a)}~Order parameter in the isotropic phase, without angular noise.
  $N=1000$ (circles) and $4000$ (crosses). From right to left: theory at
  $\rho\to0$, $\rho=10^{-3}$, $0.01$, $0.02$, $0.05$, $0.1$, $0.2$, $0.3$.
  \textbf{(b)}~Dependence of the isotropic--polar transition on the density,
  with no angular noise. \textbf{(c)}~Transition lines in the $(\alpha,
  D/\lambda)$-plane, at fixed density~$\rho=0.01$. Solid lines are theoretical
  results at $\tau_v=\infty$. Upward triangles~(red) and downward
  triangles~(blue) are transitions measured numerically by respectively
  increasing and decreasing~$D/\lambda$ quasi-statically. \textbf{(d)}~Order
  parameter obtained by increasing (red) and decreasing (blue) the angular
  noise.}
\end{figure}%
\subsection{Molecular dynamics (MD) simulations}
We now establish the phase behaviour of the model for $N$~particles.
MD~simulations were performed at $\tau_v=4$ with $N=1000$ or $N=4000$,
focusing on the dilute regime $\rho\ll1$ (see below for a discussion of
the effect of~$\tau_v$). We are thus left with two microscopic
parameters, namely $\alpha$ and~$D/\lambda$. Also, the system size is
chosen not too large, in order to keep the system spatially homogeneous,
which we have checked by visual inspection. We measured the order
parameter $\psi(t)=\bigl|\sum_i \bfv_i(t)\bigr|/N$, which is of
order~$1/\sqrt{N}$ for the isotropic state and of order one for the
polar state.

Let us first look at the case without angular noise, $D/\lambda=0$. We
initialized simulations from random isotropic conditions and waited for
the isotropic state to eventually destabilize. When a stationary state
was reached, we started to average the order parameter over
time,~$\langle\psi\rangle$. As shown in Fig.~\ref{fig:1}c, we found the
isotropic state to be stable at low values of~$\alpha$, whereas it
becomes unstable at larger values, in favour of a polar state. Between
the two phases, an abrupt discontinuous transition takes place
at~$\alpha_*$. Quite remarkably, in the whole polar phase the dynamics
converges to $\psi=1$, where particles are all strictly parallel.
Further, choosing some random state with $\psi\approx1$ as initial
condition, we found that the polar state $\psi=1$ is stable for all
$\alpha>0$, in particular also when $\alpha<\alpha_*$.

In Fig.~\ref{fig:transition}a, we show again the (now rescaled) order
parameter in the isotropic state, this time for different densities. For
a given density, the data for different values of~$N$ collapse, showing
that finite-size effects are under control. In all cases we observed
convergence to the fully polar state beyond the points shown. The
theoretical framework used below adds the line for $\rho\to0$, drawn in
black. Increasing density here clearly favours the polar state. The
departure of the transition due to density effects,
$1-\alpha_*(\rho)/\alpha_*(0)$, is plotted in Fig.~\ref{fig:transition}b.

Adding angular noise to the trajectories quite changes the picture.
During simulations we first increased~$D/\lambda$ quasi-statically and
then decreased it again. The transition was measured by looking at the
maximum of the fluctuations of the order parameter among many
realizations of the dynamics.  The resulting phase diagram at density
$\rho=10^{-2}$ is shown in Fig.~\ref{fig:transition}c. In agreement with
intuition,the isotropic state is always stable at strong enough angular
noise. When decreasing the noise we pass into the polar phase, but the
nature of this transition can be either discontinuous or continuous. For
values $\alpha<\alpha_c\approx0.157$ the transition has some hysteresis,
as indicated in the upper panel of Fig.~\ref{fig:transition}d. The
discontinuous nature of the transition is thus robust when adding
angular noise, and the phase areas in Fig.~\ref{fig:transition}c
overlap, presenting an area which we could call coexistence region if
the system were not homogeneous. For $\alpha>\alpha_c$, the hysteresis
is no longer observed at our level of numerical precision%
\footnote{In the last three points of Fig.~\ref{fig:transition}c, for
    $\alpha\ge10$ there is a slight hysteresis which disappears for slower
    annealing rate of $D/\lambda$.
},
as can be seen in the lower panel of Fig.~\ref{fig:transition}d.  At
$\alpha_c$, the coexistence zone vanishes into a single line of
transition (tricritical point~\cite{ChaLub95, Baglietto2013,
Romenskyy2014}).  Interestingly, once in the polar phase, further
increasing $\alpha$ leads to a re-entrant transition towards the
isotropic phase. 

\subsection{Kinetic theory framework}
We now rationalize these numerical observations in the context of
kinetic theory, using the properties of the binary scattering, following
Refs.~\cite{ Hanke2013, pdp }. We summarize here the procedure
described in Ref.~\cite{pdp}. In the dilute regime, the mean free-flight
time $\lambda^{-1}$ is long enough so that particles have mostly reached
their stationary velocity~$v_0$ before interacting with another particle
($\tau_{v},\tau_n \ll \lambda^{-1}$).  The binary scattering of
self-propelled particles does not conserve momentum, and that is why a
polar state can emerge from an isotropic initial condition.  Assuming
molecular chaos and choosing the von~Mises distribution as an ansatz for
the angular distribution of the velocities, one can write down an
evolution equation for the order parameter~$\psi$. This equation can
then be expanded, up to order~$\psi^3$ to study the stability of the
isotropic phase~\cite{pdp}:
\begin{equation}
  \label{eq:dpsidt}
  \frac1\lambda \frac{\text{d}\psi}{\text{d}t}
  \simeq \bigl(\mu - D/\lambda\bigr)\psi - \xi \psi^3,
\end{equation}
where the coefficients are given by
\begin{align}
  \mu &:= \bigl\langle \pdp \bigr\rangle_0 , \label{eq:mu} \\
  \xi &:= \bigl\langle (\tfrac12 - \cos\Delta)\: \pdp \bigr\rangle_0, \label{eq:xi} \\
  \label{eq:avg0}
  \langle f \rangle_0 &:= \frac14 \int_{-1}^1 \text{d}b \int_0^\pi\text{d}\Delta\: \Bigl|\sin
\frac
\Delta2\Bigr| f(b,\Delta).
\end{align}
\begin{figure*}
  \centering
  \includegraphics{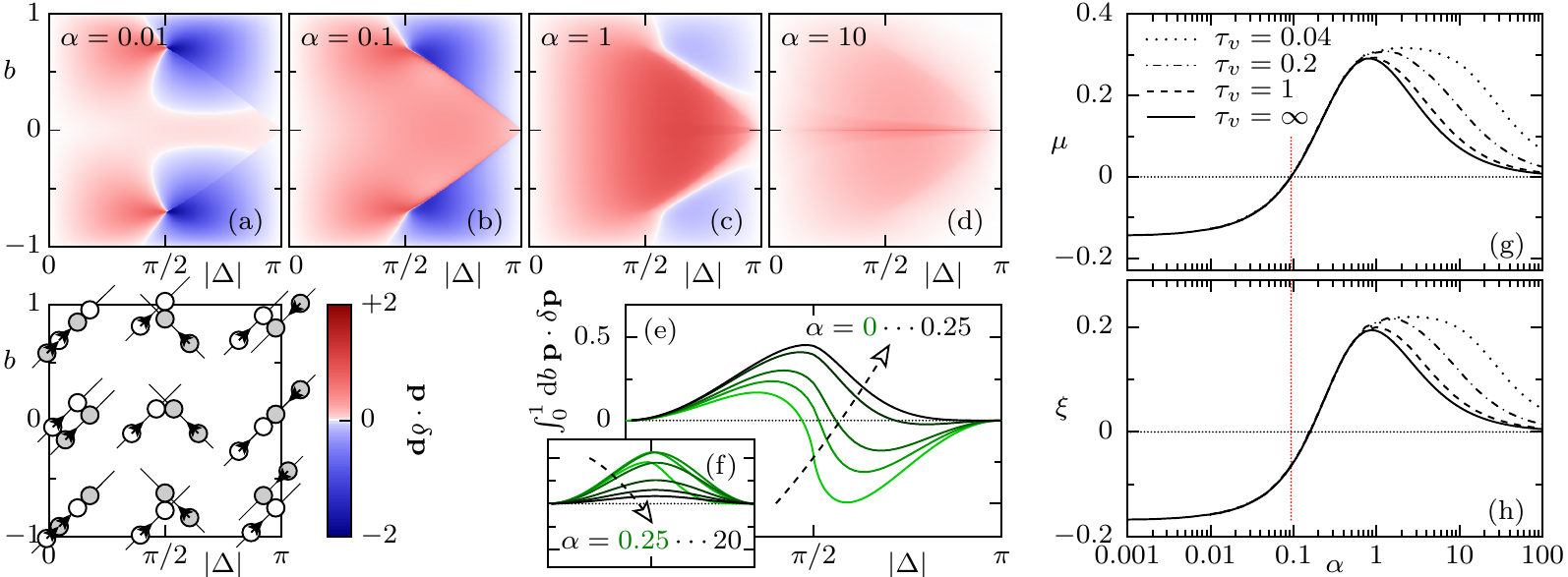}%
  \caption{\label{fig:scattering}%
  \textbf{Top:}~Colour maps of the full scattering function $\pdp(b,\Delta)$.
  The geometry of the collisions and the colour scale are shown in the
  bottom-left panel. \textbf{Bottom--middle:}~the partially integrated
  scattering function as a function of the incoming angle~$\Delta$, for
  different values~$\alpha$, $\tau_v=4$. \textbf{Right:}~Fully integrated
  scattering functions~$\mu$ and~$\xi$, defined in Eqs.~\eqref{eq:mu}
  and~\eqref{eq:xi} plotted as a function of $\alpha$ for different $\tau_v$.
  The red vertical line indicates the transition in the absence of noise.}
\end{figure*}
In the stationary state, the left-hand size of Eq.~\eqref{eq:dpsidt}
vanishes.  The transition line is obtained by solving the equation
$\mu(\alpha_*)=D/\lambda$ for $\alpha_*$, while the sign
of~$\xi(\alpha_*)$ at the transition tells whether it is continuous or
discontinuous. The coefficient $\mu$ is exact within the assumptions of
kinetic theory, while $\xi$~should depend on the ansatz used for the
angular distribution. Both coefficients are an average over all
pre-scattering parameters, as given in Eq.~\eqref{eq:avg0}, where $b$~is
the impact parameter and $\Delta$~is the angle between the incoming
particles' velocities. The averaging needs not be done over the norms of
the velocities, since those are fixed to $v_0=1$. We have checked
explicitly that this assumption holds very well in the numerical
simulations in the dilute regime.
In Eqs.~\eqref{eq:mu} and~\eqref{eq:xi}, $\bfp$~is the pre-scattering momentum
of the two colliding particles, and $\delta\bfp$~is the change of their
momentum by the scattering event. 
In contrast with Ref.~\cite{ Hanke2013 }, the above equations explicitly identify
the \emph{forward component of the momentum change} $\pdp$, as the proper
quantity to describe the alignment in a binary scattering event.
The predictions thus depend only on the microscopic
details of the deterministic dynamics through the scattering
function~$\pdp(b,\Delta)$, which we now explore for the
model~\eqref{eq:dyn}.

\subsection{Binary scattering}
Remind that a scattering event can consist of several, if not many,
hard-disk collisions, such as depicted in Fig.~\ref{fig:1}b. Before the
first collision, both particles are taken to be in the stationary state,
i.e.~$\bfv=\bfn$. After the collision, we integrate Eqs.~\eqref{eq:dyn}
numerically until another collision possibly occurs. The binary
scattering is considered over when both particles have again reached
their stationary state and are heading away from each other.  Repeating
the procedure in the whole range of initial parameters $(b,\Delta)$
yields the scattering function $\pdp(b,\Delta)$.
Figure~\ref{fig:scattering} shows this function for several~$\alpha$, as
well as its integral over~$b$ and the full integrals yielding the
coefficients $\mu$ and~$\xi$.

Let us stress that $\pdp$~is not changed by the collision itself, which
conserves momentum. All (dis)alignment must here come from the
relaxation of the post-collisional value of $|\bfv|$ to unity.
Fig.~\ref{fig:scattering}a--d shows that scattering at low angular
separation, small~$\Delta$, always creates forward momentum. In other
words, two nearly parallel particles that interact become even more
parallel, which gives rise to an effective alignment, $\pdp>0$. On the
other hand, for small enough $\alpha$, particles that enter in
interaction frontally ($\Delta\approx\pi$) tend to disalign, except for
special symmetry such as~$b\approx0$. Increasing~$\alpha$ favours
aligning scattering events until eventually only aligning events remain.
This is best summarized by integrating out all parameter dependence
except the incoming angle, as is plotted in
Fig.~\ref{fig:scattering}e,f.

The coefficients $\mu$~and $\xi$ are then obtained by integrating
over~$\Delta$, with weights prescribed by Eqs.~\eqref{eq:mu}
and~\eqref{eq:xi}.  Their dependences on the microscopic parameters of
the dynamics, $\alpha$~and $\tau_v$, are shown in
Fig.~\ref{fig:scattering}g,h.  In the absence of noise, the transition
occurs for $\alpha=\alpha_*$ such that $\mu(\alpha_*)=0$;
$\xi(\alpha_*)$ is negative, hence the discontinuous transition. When
angular noise is added, the transition is obtained by solving the
equation $\mu(\alpha_*)=D/\lambda$. From the shape of the curve
$\mu(\alpha)$, one obtains two values $\alpha_{\pm}$, with $\alpha_- \to
\alpha_*$ and $\alpha_+\to\infty$ when $D\to0$.  As $D/\lambda$ is
increased, $\xi(\alpha_-)$ eventually becomes positive and the
transition turns continuous at a tricritical point
$(\alpha_c,D_c/\lambda)$.  Note that $\xi(\alpha_+) >0$: the re-entrant
transition is always continuous.  Regarding the role of $\tau_v$,
$\alpha_-$ is practically independant on $\tau_v$, while $\alpha_+$
increases when $\tau_v$ decreases.
The theoretical predictions are shown as solid lines in
Fig.~\ref{fig:transition}c. The agreement with the MD simulations data
for density $\rho=10^{-2}$ is excellent. The small shift of the measured
transition lines to the left with respect to the theoretical one comes
from finite-density effects.

Finally, we also learn from the  examination of the scattering maps
that, in the absence of noise, the polar phase $\psi = 1$ is actually an
absorbing phase~\cite{Hinrichsen2000}: this is because \emph{all} binary
scattering events at small~$\Delta$ have $\pdp>0$. When all particles in
a system are sufficiently parallel, binary scattering events can only
align the system more. This is true for all~$\alpha$, and most
remarkably for $\alpha\to0$.

Altogether, our kinetic theory description, using the von Mises ansatz
for the angular distribution, captures quantitatively all the
phenomenology reported in the numerical simulations at low enough
density. It however relies on the numerical evaluation of the scattering
maps. In the last part of the paper, we would like to provide some
intuition on the origin of the peculiar form of these maps. Also, we
will elucidate the role of the multiple collisions which can take place
during a scattering event.

%
\begin{figure}
  \centering
  \includegraphics{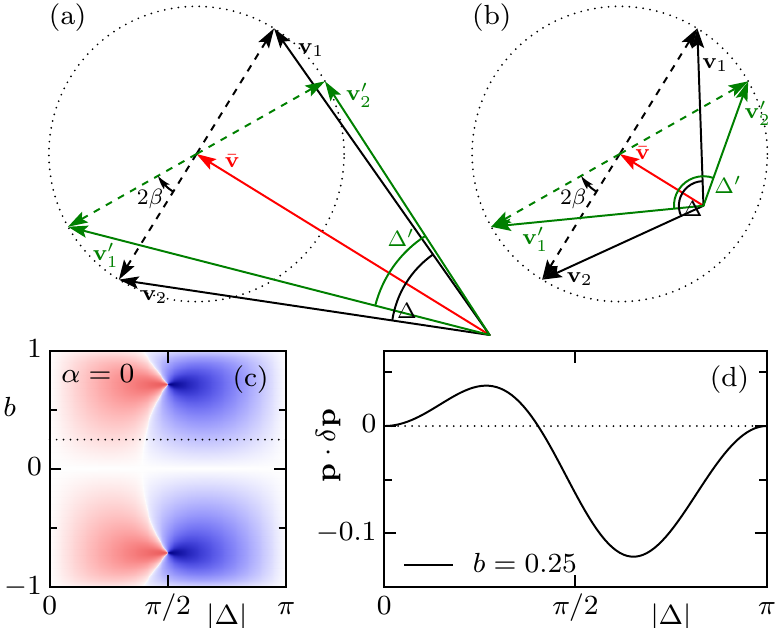}%
  \caption{\label{fig:collision}%
  \textbf{(a,b)}~Elastic collisions are rotations by an
  angle~$2\beta=2\asin(b)$ in the reference frame of the center-of-mass
  velocity~$\bar\bfv$. Primed quantities (green) are post-collisional, black
  ones are pre-collisional. \textbf{(c)}~The scattering function~$\pdp$
  resulting from a single collision. \textbf{(d)}~A cut through~(c) for the
  value of~$b$ used in~(a,b).}
\end{figure}%
\subsection{Analytical limits}
When we take the limit~$\tau_n\to0$, the vector~$\bfn$ has no
persistence at all. After two particles collide elastically, the $\bfn$
rotate to their respective $\bfv$ instantaneously, the two particles
cannot collide a second time and the post-collisional relaxation of
$|\bfv|$ to unity occurs on a straight trajectory. We take advantage of
this simplification to compute the scattering map in the case
$\alpha=0$. Because velocities have equal modulus, we can use the nice
visualisation of an elastic collision as a rotation in the
center-of-mass frame (Fig.~\ref{fig:collision}a,b). One can then find an
analytic expression\footnote{%
    $\pdp(b,\Delta) = 2\cos\frac{\Delta}{2} \sum\limits_{s=-1,1}
    \biggl[\frac{\cos\tfrac{\Delta}{2} +
    2sb\sqrt{1-b^2}\sin\tfrac{\Delta}{2}}{\sqrt{1 + 2sb\sqrt{1-b^2}\sin\Delta}} -
    \cos\frac{\Delta}{2}\biggr]$
}
for $\pdp(b,\Delta)$, which is plotted in Fig.~\ref{fig:collision}c.
This plot nicely completes the series of varying~$\alpha$ from
Fig.~\ref{fig:scattering}a-d. The essential structure of the scattering
maps can be accounted for by the sole ingredients present in the case
$\alpha = 0$. Inelastic collisions and persistence of~$\bfn$ are not
essential.  Conversely, non-conservation of momentum due to the
relaxation of the particle speeds to unity is crucial.

\begin{figure}
  \centering
  \includegraphics{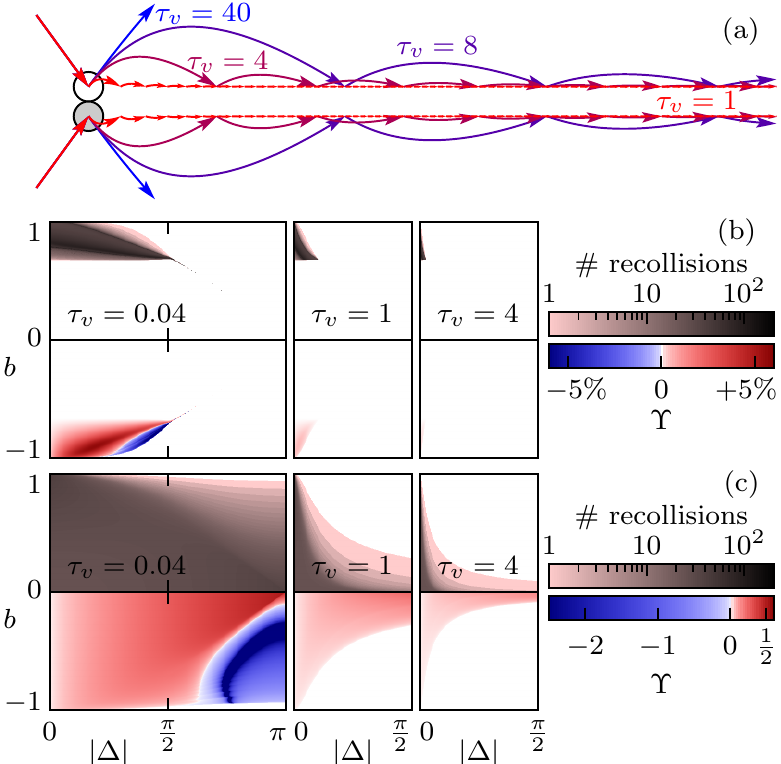}%
  \caption{\label{fig:recoll}
  Recollisions.
  \textbf{(a)}~Symmetric binary scatterings for $\alpha=5$.
  \textbf{(b, c)}~Number of recollisions in each scattering (top rows)
  and their relative contribution to $\pdp$ (bottom rows).
  $\alpha=0.1$ for (b), $\alpha=10$ for (c).
}
\end{figure}

At values $\alpha>0$, the vector~$\bfn$ has non-vanishing persistence,
which results in curved trajectories and possible recollisions. This
could lead to the effective alignment mechanism which was proposed to be
at the root of the collective motion in the experiment of vibrated polar
disks~\cite{ Deseigne2010, Deseigne2012, Weber2013 }.
Concerning the influence of recollisions in the scattering, we ask
whether they contribute significantly to~$\mu$ and $\xi$. In
particular, how numerous are they depending on the scattering geometry
$(b,\Delta)$, what is their statistical weight and how much do they
affect the scattering map $\pdp(b,\Delta)$? 
One can get an intuitive picture of the recollision mechanism looking at
a simple case, the symmetric ($b=0$) binary scatterings in
Fig.~\ref{fig:recoll}a. A~first collision makes $\bfv$~rotate from being
equal to~$\bfn$ to pointing away from the other particle. On the
following trajectory, $\bfv$~and $\bfn$~relax towards each other, on
their respective timescales~$\tau_v$ and $\tau_n$.  For small enough
$\tau_v$, the persistence of $\bfn$ allows for a recollision to take
place, and both vectors get closer to each other from one collision to
the next.
The highly symmetric case depicted in Fig.~\ref{fig:recoll}a is however
so degenerate that it has either one or infinite collisions%
\footnote{In the symmetric case, linearizing
    the dynamics around the stationary point provides analytical results about this
    mapping from one collision to the next.
    For $\alpha<1$, there is no recollision.
    For $\alpha>1$, an infinite number of recollisions occur,
    with a collision rate that converges to a positive constant ($\alpha<2$)
    or diverges ($\alpha>2$).
}.
Unfortunately, the analytic treatment of asymmetric collisions is much
too technical to gain physical insight from it and we prefer coming back
to the numerical sampling of the scattering events.

For $\alpha=\alpha_-$, as shown in Fig.~\ref{fig:recoll}b (top row),
recollisions can be numerous, but take place only in a small region of
the parameter space $(b,\Delta)$, the smaller the larger $\tau_v$.
We measure their relative contribution to $\pdp$ by $\Upsilon=\tfrac12
\sin(\Delta/2) (\pdp - \pdp|_1)/|\pdp|$, where $\pdp|_1$ is the
contribution of the first collision only.
As shown in Fig.~\ref{fig:recoll}b (bottom row), even for the smallest
value $\tau_v=0.04$, this contribution is locally less than 5~\%.  For
larger $\tau_v$, the contribution is vanishingly smaller.  Thus,
recollisions has no practical influence on the transition at
$\alpha=\alpha_-$.
Conversely, as shown in Fig.~\ref{fig:recoll}c (top row), for
$\alpha=\alpha_+$, the number of recollisions is not so large, but they
occur for wider ranges of scattering parameters and their relative
contribution to $\pdp$ is significant, see Fig.~\ref{fig:recoll}c
(bottom row).  This is all the more true for smaller $\tau_v$ and is
responsible for the dependance of $\mu$ and $\xi$ on $\tau_v$ in this
regime.

\subsection{Discussion}
Let us recast our main findings and the understanding we obtained from
the above analysis. A~homogeneous system of self propelled hard disks,
obeying the minimal deterministic Eqs.~(\ref{eq:dyn}), exhibits a sharp
discontinuous transition from an isotropic to an absorbing polar
collective motion state, when increasing the persistence of the body
axis vector $\bfn$. Adding angular noise, the transition becomes
continuous via a tricritical point. At fixed noise level, increasing
further the persistence of the body axis vector, a re-entrant continuous
transition from the polar state to the isotropic state occurs. 

All these macroscopic behaviours can be understood by the study of the
scattering maps~$\pdp$, which describe the level of alignment of all
possible scattering events. The most striking feature of these maps is
that scattering at low angular separation always creates forward
momentum. This is even true for $\alpha = 0$, for which, we could prove
it explicitly from a mechanical analysis of the collisions. This feature
explains the stability of the polar phase in the absence of noise. For
small~$\alpha$, the dis-alignment, present in the frontal collisions is
strong enough to stabilize the isotropic state.  For $\alpha>\alpha_*$,
this does not hold anymore, and the isotropic state becomes unstable.
The coexistence of stability of the two states for $\alpha<\alpha_*$
guarantees the transition to be discontinuous.

The effect of noise is most easily understood from the shape of the
curves $\mu(\alpha)$ and~$\xi(\alpha)$. The key point here is that in
the present scheme of approximation -- Boltzmann equation plus von Mises
Ansatz -- the computation of~$\xi$ is identical to that of $\mu$, apart
from the additional asymmetric factor $(1/2-\cos \Delta)$.  This does
not rely on the specificity of the microscopic model.
As a result, the curve~$\xi(\alpha)$ is similar to the
curve~$\mu(\alpha)$, with a shift towards larger~$\alpha$.
Adding noise simply shifts the transition towards larger~$\alpha$: at
some point $\xi$~changes signs at the transition, which becomes
continuous, hence the tricritical point.

Finally, the re-entrant transition from polar back to isotropic at
large~$\alpha$ (while increasing~$\alpha$) is a direct consequence of
the non-monotonous shape of~$\mu(\alpha)$. At moderate~$\alpha$, the
effective alignment~$\mu$ increases with the persistence of~$\bfn$.
However, a too large persistence reduces the alignment, because the
scattering event does not reorient~$\bfn$ significantly, and no memory
of the collision is kept: $\bfv$~relaxes to practically the same~$\bfn$
as before the scattering.  Some alignment still takes place but is most
easily destroyed by small amounts of noise. Interestingly, while the
recollisions play no role at the first transition from the isotropic to
the polar state, they here increase the effective alignement and thereby
postpone the re-entrant transition to larger~$\alpha$.

As can be seen from the above discussion, the results do not depend on
the details of the scattering maps, as long as binary scattering at low
angles produces effective alignment. The partially integrated scattering
function (Fig.~\ref{fig:scattering}e,f) is then positive for small
angles. Models which fall into this class are (i)~the here described
hard disks with~$\bfv$ and~$\bfn$, (ii)~the inelastically colliding hard
disks with only~$\bfv$~\cite{pdp}, and (iii)~the soft disks described in
Ref.~\cite{Hanke2013}. All three are self-propelled in the sense that
their free-flight speed relaxes to~$v_0$.

It is important to note that the distinctive behaviour of the above
class of system is markedly different from the original Vicsek model, in
which binary scattering at low angle effectively disaligns. In the Vicsek
model, the partially integrated scattering function~\cite[Fig.~2a]{pdp} behaves
precisely in the opposite way of the one here. It is therefore unlikely that a
coarse-graining of self-propelled hard disks (or any of the above three models)
would yield the Vicsek model. Altogether, the present new class of models,
including the self-propelled hard particles, could well play for the theory of
``simple active liquids'' the role that hard spheres play for the statistical
mechanics of gases.

\vspace{-4mm}

\bibliographystyle{eplbib}
\bibliography{refs}
\end{document}